\documentclass[12pt]{report}
\usepackage{amsmath, amssymb, graphics}
\newcommand{\mathsym}[1]{{}}

\newcommand{\be}{\begin{equation}}
\newcommand{\ee}{\end{equation}}
\newcommand{\bc}{\begin{center}}
\newcommand{\ec}{\end{center}}
\newcommand{\bea}{\begin{eqnarray}}
\newcommand{\eea}{\end{eqnarray}}

\newcommand{\pr}{\prime}

\newcommand{\nn}{\nonumber}
\newcommand{\rf}{\ref}

\newcommand{\ep}{\epsilon}

\newcommand{\noi}{\noindent}

\newcommand{\al}{\alpha}
\newcommand{\ga}{\gamma}
\newcommand{\bt}{\beta}

\newcommand{\si}{\sigma}

\newcommand{\lb}{\label}

\newcommand{\qu}{\quad}

\usepackage{amsmath, amssymb, graphics}

\begin{document}
\bc
{\large{\bf{Symmetry Analysis for a Generalized Kadomtsev-Petviashvili Equation}}}

B. Mayil Vaganan$^1$, D. Pandiaraja$^2$ and M. Senthilkumaran$^2$\\

$^1$Department of Applied Mathematics and Statistics, 
Madurai Kamaraj University, Madurai-625021, India\\
$^2$Department of Mathematics, Thiagarajar College, Madurai-625009, India\\ 
\ec

\noi {\bf Abstract}

 A generalized Kadomtsev-Petviashvili equation (GKPE)
$(u_t+u u_x + \beta(t)u +\gamma(t)u_{xxx})_x+\sigma(t)u_{yy}\ = \ 0$ is shown to admit 
an infinite-dimensional Lie group of symmetries  when $\bt(t),\ \ga(t)$ and $\si(t)$ are arbitrary.  The Lie algebra of this symmetry group contains two arbitrary functions $f(t)$ and $g(t)$. Further, 
low-dimensional subalgebras and  physically meaningful five dimensional Lie algebra containing translation and Galilei transformation are derived.  A solution of GKPE involving two arbitrary functions of time $t$, in addition to $f(t)$ and $g(t)$, is obtained using an one-dimensional subalgebra.\\


AMS Classification Numbers: 22E60, 27E70, 34A05, 35G20.

{\bf 1. \ Introduction}
 
Kadomtsev-Petviashvili (KP) equation
\be
(u_t+\frac{3}{2}u u_x + \frac{1}{4}u_{xxx})_x+\frac{3}{4}u_{yy}\ =\ 0, \lb{I1}
\ee
known also as the two-dimensional Korteweg-de Vries equation  arises in the study of long gravity waves in a single layer, or multilayered shallow fluid, when the waves propagate predominantly in one direction with a small perturbation in the perpendicular direction.  The mathematical interest of  KP equation stems from the fact that it is associated with an infinite-dimensional Lie groups.  It is integrable in the sense of allowing Lax pair,  conservation laws, solitons,  and periodic solutions (See [3] and references 1-11 in [3]). 

A prototype example of the derivation of a generalized KP (GKP) equation from Euler equations in somewhat realistic conditions was given by David, Levi and Winternitz [5]. David, Levi and  Winternitz [4] studied the symmetries and reductions for a  generalized KP  equation
\be
(u_t+u u_x +u_{xxx})_x+\sigma(t)u_{yy}\ =\ 0. \lb{Ia1}
\ee

Brugarino and Greco [2] studied   VCKP equation 
\be
(u_t+ a(x, y, t)u + b(x, y, t)u_x + f(x, y, t)u   u_x  +g(x, y, t)u_{xxx})_x+h(x, y,t)u_{yy}  = k(x, y, t),\lb{Ij2}
\ee
to determine the conditions on the coefficient functions under which (\ref{Ij2}) passes the Painlev\'e test.

  G\"ung\"or and Winternitz [9] classified another VCKP equation
\be
(u_t+f(x, y, t)u u_x  +g(x, y, t)u_{xxx})_x+h(x, y,t)u_{yy}  = 0,\lb{Ij1}
\ee
into equivalence classes under fibre preserving point  transformations with a nonzero Jacobian.

G\"ung\"or and Winternitz [10], using the allowed transformation, transformed yet another VCKP equation
\be
(u_t+p(t)u u_x  +q(t)u_{xxx})_x+\sigma(y,t)u_{yy}+a(y, t) u_y + b(y, t) u_{xy} + c(y, t) u_{xx} + e(y, t) u_x + f(y, t) u + h(y, t) = 0,\lb{Ij3}
\ee
into the canonical form 
\be
(u_t+u u_x  +u_{xxx})_x+ \epsilon u_{yy}+a(y, t) u_y + b(y, t) u_{xy} + c(y, t) u_{xx}    + f(y, t) u  = 0,\qu \epsilon = \pm 1,\lb{Ij4}
\ee
and investigated its  group theoretical properties in order to establish the conditions on the coefficient functions $a, b, c$ and $f$ under which (\ref{Ij3}) admits an infinite-dimensional symmetry group having a Kac-Moody-Virasoro structure.

Here If we  consider a  GKPE  
\be
(u_t+\alpha^\pr(t)u u_x + \beta^\pr(t)u +\gamma^\pr(t)u_{xxx})_x+\sigma^\pr(t)u_{yy}\ =\ 0,\ \  \gamma^\pr(t),\sigma^\pr(t)\neq 0. \lb{I0}
\ee
The point transformation 
\be
\overline{t} = \int_{t_0}^t\ \al(s)ds, 
\ee
 replaces (\ref{I0}) by an equation of the form  
\be
(u_t+u u_x + \beta(t)u +\gamma(t)u_{xxx})_x+\sigma(t)u_{yy}\ =\ 0,\ \  \gamma(t),\sigma(t)\neq 0. \lb{I2}
\ee
Equation (\rf{Ia1}) is a special case of (\rf{I2}) when $\beta(t)= 0$ and $\gamma(t)= 1$. 
In this paper we study the symmetry properties of the GKPE (\rf{I2}) by 
closely following the works of David, Kamran, Levi and Winternitz [3] and G\"ung\"or [7-8].
To be precise, we shall  show that the  GKPE (\ref{I2})
admits an  infinite-dimensional symmetry group and determine the corresponding Lie algebra,  extend it by specifying the coefficient functions $ \beta(t), \gamma(t), \sigma(t)$, and classify the one- and two-dimensional subalgebras of the symmetry algebra  under the adjoint action of the symmetry group in order to reduce (\rf{I2}) to (1+1)-dimensional partial differential equations (PDEs) and then to ordinary differential equations (ODEs). The symmetry algebra is found to involve two arbitrary functions $f(t)$ and $g(t)$. It is shown that (\ref{I2}) reduces to a linear PDE $W_{yy}(y, t) = F(f(t), f^\prime(t))$ and also to a VCKdVE (\rf{I41}).   Several symbolic manipulation packages are available  for calculating the symmetry group of PDEs (See Yao Ruo-Xia and Lou Sen-Yue [14] and references therein).  In this work we use MathLie [6] to determine the symmetry group of  GKPE (\ref{I2}).

This paper is organised as follows: In section 2	we derive the symmetry group and study  the structure of the symmetry algebra of the GKPE (\rf{I2}).  Section 3	is devoted to the   determination of physically interesting finite-dimensional algebra by restricting $f(t)$ and  $g(t)$ to first degree polynomials. In section 4	we give the classification of low-dimensional subalgebras of the GKPE algebra, namely those of dimension $n=1, 2$ into conjugacy classes under the adjoint action of the symmetry group of the GKPE (\rf{I2}).  This is done mainly to elucidate the structure of the considered infinite-dimensional Lie algebra and to establish the applicability of tools developed for classifying subalgebras of finite-dimensional Lie algebras. 
In section 5 we  reduce the GKPE (\ref{I2}) into (1+1)-dimensional PDEs using   the one-dimensional subalgebras of GKPE algebra. In section 6 we use two isomorphy classes of two-dimensional algebras, namely, Abelian and non-Abelian, to reduce the PDEs obtained in section 5 to ODEs. In section 7 we write down the general form of the reduced ODEs and are transformed to special cases of equations introduced by   Mayil Vaganan and Senthilkumaran [11]. Finally in section 8	we summarise the results of the present work. 

{\bf 2. \ The symmetry group and  Lie algebra of the GKPE (\rf{I2})}

If (\rf{I2}) is assumed to be invariant under Lie group of infinitesimal transformations (Olver [11], Bluman and Kumei [12])
\be
x_i^{*} = x_i + \epsilon \xi_i(x, y, t, u) + O(\epsilon^2),\qu i=1,2,3,4,
\ee
where $\xi_1 =\xi, \xi_2 = \eta, \xi_3 =\tau, \xi_4=\phi$, then the corresponding vector field $V$ is
\be
V=\tau (x,y,t;u)\ \partial_t + \xi (x,y,t;u)\ \partial_x+\eta(x,y,t;u)\ \partial_y +\phi(x,y,t;u)\ \partial_u.\lb{I3}
\ee
Then the fourth prolongation of $V$ must satisfy
\be
pr^{(4)}V\Omega(x,y,t;u)|_{\Omega (x,y,t;u)=0}=0.\lb{I4}
\ee 
where $\Omega(x,y,t;u)=0$ is (\rf{I2}) and $pr^{(4)}$ stands for the fourth prolongation of the vector field $V$.
The defining equations are obtained from (\ref{I4}) and solved for the infinitesimals $\xi, \eta, \tau, \phi$ for the following five cases:

{\bf Case i.}\ $\beta, \gamma, \si$ are arbitrary.

The infinitesimals $\xi, \eta, \tau $ and $\phi$ are obtained as
\be
\xi = f-\frac{y}{2}\left(\frac{g^{\prime}}{\si}\right),\qu 
\eta = g, \qu \tau = 0, \qu
\phi = f^{\prime}-\frac{y}{2}\left(\frac{g^{\prime}}{\si}\right)^{\prime}.\lb{I8}
\ee
The symmetry algebra of (\rf{I2}) is an infinite-dimensional Lie algebra $L_p = \left\{V \right\}$, where
\bea
V&=&X(f)+Y(g),\lb{I9}\\
X(f) &=& f\partial_x + f^{\prime}\partial_u,\lb{I10}\\
Y(g) &=& -\frac{y}{2}\left(\frac{g^{\prime}}{\si}\right)\partial_x + g\partial_y -\frac{y}{2}\left(\frac{g^{\prime}}{\si}\right)^{\prime} \partial_u . \lb{I11}
\eea
Here $f(t)$ and $g(t)$ are arbitrary smooth function and satisfy commutation relations
\bea
[X(f_1), X(f_2)]=0, \quad [X(f), Y(g)]=0, \quad [Y(g_1), Y(g_2)]= X\left[\frac{1}{2\sigma}\left(g_2g_1^{\prime}-g_1 g_2^{\prime}\right)\right].\lb{I12}
\eea

As $\partial_t$ does not appear in $V$, the Lie algebra $L_p$ is not of Virasoro type (cf. G\"ung\"or [7]).  Each of the vector fields $X(f)$ and $Y(g)$ can be integrated separately to obtain the Lie group of transformations. Thus if $u(x,y,t)$ is any solution to (\rf{I2}),  then so are
\bea
u^{\prime}(x^{\prime},y^{\prime},t^{\prime}) &=& u(x-\epsilon f(t), y, t)+ \epsilon f^{\prime}(t),\lb{I13}\\
u^{\prime}(x^{\prime},y^{\prime},t^{\prime}) &=& u\left(x-\frac{g^{\prime}}{2\si}y \epsilon-\frac{gg^{\prime}}{4\si}\epsilon^2, y+g \epsilon, t\right)  -\frac{1}{2}\left(\frac{g^{\prime}}{\sigma}\right)^{\prime}\left(y\epsilon +g \frac{\epsilon^2}{2}\right).\lb{I14}
\eea

Now we shall show that the algebra $L_p$ becomes larger when we specify the functions $\beta, \gamma, \si$.  We list below 3 such extensions of $L_p$.  In the foregoing analysis $c_1, \ lambda \in R$.

{\bf Case ii.}\ $\beta(t)=\beta, \gamma(t)=\gamma, \si(t)=\si $, where $ \beta, \gamma, \si$ are constants.

It is found that $\tau$ is no longer zero, but is given by $\tau=c_1$.  Therefore, in this case, the symmetry algebra $L_1$ is represented by  (\rf{I9}) and  $T_0=\partial_t$.

Now the Lie algebra $L_1$ with the basis $X(f), Y(g)$ and $T_0$ can be written as a semidirect sum
\be
L_1=\left\{ X(f), Y(g)\right\} \oplus_s \left\{T_0 \right\}.\nn
\ee

{\bf Case iii.}\ $\beta, \gamma$ are constants and $\si(t)= e^{\lambda t}$

The infinitesimals which undergo changes are $\eta$ and $\tau$.  Indeed, we find that
\be
\eta = c_1\ y\ + g(t) \quad  {\rm and}\qu \tau = \frac{2}{\lambda}c_1 . \lb{I15}
\ee

The Lie algebra $L_2$ has an additional generator
\be
D_{\lambda} =\frac{\lambda}{2}\ y\ \partial_y + \partial_t, \lb{I16}
\ee
which is a scaling in the $y$-direction and translation in time $t$.  Thus the basis of $L_2$ is $X(f), Y(g)$ and $D_{\lambda}$.  In this case we may write $L_2$ as
\be
L_2=\left\{X(f), Y(g) \right\} \  \oplus_s \ \left\{ D_{\lambda}\right\}.\nn
\ee

{\bf Case iv.} \ $\beta, \si$ are constants and $\gamma(t)= e^{\lambda t}$.

Here the infinitesimals are
\be
\xi = f+ \frac{c_1}{2\beta}x -\frac{1}{2\si}yg^{\prime}, \qu 
\eta = \frac{c_1 y}{4\beta}+g,\qu
\tau = \frac{3c_1}{2\beta \lambda},\qu
\phi= \frac{c_1 u}{2\beta}+f^{\prime}-\frac{y g^{\prime \prime}}{2\si}.\lb{I22}
\ee

Hence the basis of the Lie algebra $L_3$ is now given by the three generators
\be 
X(f), Y(g)\quad {\rm and} \quad E_{\lambda} = \frac{\lambda}{3}x\partial_x+\frac{\lambda}{6}y\partial_y+\partial_t+\frac{\beta\lambda}{3}u\partial_u.\lb{I23}
\ee
The gererator $E_\lambda$ contains scalings in $x,y$ and $u$ directions and translation in $t$.
We write the Lie algebra $L_3$ as 
\be
L_3=\left\{X(f), Y(g) \right\} \  \oplus_s \ \left\{E_{\lambda} \right\}.\nn
\ee
It is now easy to infer the following facts:

(i) When $\beta, \gamma$ and $\si$ are arbitrary functions of time $t$, the Lie algebra  $L_p=\left\{X(f), Y(g) \right\} $, is of infinite-dimensional with the basis given by two generators $X(f), Y(g)$.

(ii) If we restrict $\beta, \gamma$ and $\si$ to constants then the Lie algebra $L_p$ gets enlarged to $L_1$ as $L_1$ is found to be the semi-direct sum of $L_p$ and $T_0$.

(iii) If we only take $\beta, \gamma$  to be constants and $\si(t)= e^{\lambda t}$, then Lie algebra $L_2$, in addition to $X(f)$ and $Y(g)$, contain another basis element  $D_{\lambda}$. 

(iv) If $\gamma(t)=e^{\lambda t} \, $ and $\beta, \si$ are taken as constants, then  Lie algebra $L_3$ is shown to be generated by the three infinitesimal generators $X(f), Y(g)$,  and  $E_{\lambda}$.

The commutator table amongst $X(f), Y(g), T_0, D_\lambda, E_\lambda$ is given below:

\bc
\begin{tabular}{|l|l|l|l|l|l|}
\hline
	 &$	X(f)$	& $	Y(g)$	& $	T_0$	& $	D_\lambda$ 	& $	E_\lambda$\\
\hline
$X(f)$&	0&$0$&$-X(f^\prime)$&$-X(f^\prime)$&$X(\frac{\lambda}{3}f-f^\prime)$\\
\hline
$Y(g)$&	0&$0$&$-Y(g^\prime)$&$X(\frac{\lambda}{2}\frac{yg^\prime}{\sigma})+Y(\frac{\lambda}{2}g - g^\prime)$&$Y(\frac{\lambda}{6}g-g^\prime)$\\
\hline
$T_0$&$X(f^\prime)$&$Y(g^\prime)$&0&0&0\\
\hline
$D_\lambda$&$X(f^\prime)$&$-X(\frac{\lambda}{2}\frac{yg^\prime}{\sigma})-Y(\frac{\lambda}{2}g - g^\prime)$&0&0&0\\
\hline
$E_\lambda$&$-X(\frac{\lambda}{3}f-f^\prime)$&$-Y(\frac{\lambda}{6}g-g^\prime)$&0&0&0\\
\hline

\end{tabular}
\ec
\bc Table- 1. \ec

{\bf 3. \ A finite-dimensional subalgebra of physical transformations}

We shall now systematically classify $L_p$ into finite-dimensional subalgebras of physical interest. If we choose $f(t)=g(t)=1$ and $f(t)=g(t)=t$ respectively, then we have
\be
X(1)=\partial_x=X, \quad Y(1)=\partial_y =Y, \lb{I24}
\ee
 and
\be
X(t)=t\partial_x + \partial_u= B, \quad Y(t)= -\frac{y}{2\si} \partial_x + t \partial_y=R.\lb{I25}
\ee
Here $X$ and $Y$ are translations in $x$ and $y$ respectively and $B$ is a Galilei transformation in the $x$ direction. Finally $R$ is a combination of a Galilei transformation in the $y$ direction and a pseudo-rotation.   

Now the Lie algebra $L_0$ corresponding to the GKPE 
\be 
(u_t+u u_x + \beta u +\gamma u_{xxx})_x+\sigma u_{yy}\ =\ 0,\lb{I26} 
\ee 
where $\beta,\gamma$ and $\sigma $ are constants, is
\be
L_0=\left\{X, B, R, Y, T_0 \right\}
\ee
which is of dimension five.  The commutator table for $L_0$ is
\bc
\begin{tabular}{|l|l|l|l|l|l|}
\hline
	 &$X$	& $B$	& $R$	& $Y$ & $T_0$\\
	 \hline
$X$&	0&$0$&0&0&0\\
\hline
$B$&$0$&0&$0$&0& $-X$\\
\hline
$R$&0&0&$0$&$-\frac{X}{2\si}$ & $-Y$\\
\hline
$Y$&0&0&$\frac{X}{2\si}$& 0 & 0\\
\hline
$T_0$&0&$X$& $Y$ &0&0\\
\hline
\end{tabular}
\ec
\bc Table-2\ec

{\bf 4. \ Low-dimensional subalgebras of the symmetry algebra of GKPE (\ref{I2})}

In order to obtain the solutions of the GKPE (\rf{I2}) by symmetry reduction, it is essential to identify the low-dimensional subalgebras of the GKPE symmetry algebra.  In particular, we need to find subalgebras that correspond to Lie groups having orbits of codimension 2 or 1 in the four-dimensional space coordinated by $(x, y, t, u)$.  We therefore classify the one-dimensional subalgebras into conjugacy classes under the adjoint action of the symmetry group of the GKPE (\rf{I2}). In the foregoing analysis the results given in (\rf{I12}) and Table-1 are used.

{\bf Case 1.} \ $\beta, \gamma, \si$ - arbitrary functions of time $t$

If we take conjugation of $V=X(f)+Y(g)$ by $Y(G)$, where $G(t)$ is to be determined, then, in view of the commutation relation (\rf{I12}), we have
\bea
Ad\left\{\exp(\epsilon Y(G))\right\}V&=& V-\epsilon \ [y(G), V]\nn\\
&=& V-\epsilon \ [Y(G), X(f)+Y(g)]\nn\\
&=& V-\epsilon \ [Y(G), X(f)] -\epsilon [Y(G), Y(g)]\nn\\
&=& V- \epsilon \ X \ \left(\frac{1}{2\si}(gG^{\prime}-Gg^{\prime})\right)\nn\\
&=& X(f)+Y(g) - X \left(\frac{\epsilon}{2\si}(gG^{\prime}-Gg^{\prime})\right)\nn\\
&=& X \left(f- \frac{\epsilon }{2\si}(gG^{\prime}-Gg^{\prime})\right)+Y(g).\lb{I27}
\eea
Now we fix $G(t)$ as
\be
G(t)= 2bg(t)\int ^{t}_{1}  \frac{\sigma(t) f(t)}{[g(t)]^2}dt + c g(t),\lb{I28}
\ee
where $b$ and $c$ are arbitrary constants.  We choose $G(t)$ given by  (\rf{I28}) as the function labelling the generator $Y(G)$ of the symmetry algebra of the GKPE (\ref{I2}), and $\epsilon = b^{-1}$ as the value of the parameter $\epsilon$ of the one-parameter subgroup associated with $Y(G)$.  Then it is evident that $V$ is conjugate to $Y(g)$ if $g\neq 0$ and $V$ is conjugate to $X(f)$ if $g=0$.  Therefore it is enough to consider the two one-dimensional subalgebras namely $L_{p,1} = \left\{ X(f)\right\}$ and $L_{p,2}=\left\{Y(g)\right\}$ instead of the full symmetry algebra $L_p$ itself.

{\bf Case 2.} \ $\beta, \gamma, \sigma $ - arbitrary constants.\\ 
If we take conjugation of $V_1 = X(f)+Y(g)+aT_0 $ , $a \ne 0$ by $X(F)+Y(G)$   we obtain
\bea
Ad\left\{\exp(\ep X(F)+\delta Y(G))\right\}V_1 &=& V_1-\epsilon \ [X(F), V_1] -\delta [Y(G), V_1]\nn\\
&=& V_1- \epsilon \ [X(F), aT_0] -\delta [Y(G), Y(g)]-\delta  [Y(G),aT_0]\nn\\
&=& aT_0 + X(f+a\epsilon F^\prime - \frac{\delta}{2\sigma}[gG^\prime-Gg^\prime])+Y(g + a\delta G^\prime).\lb{I29}
\eea
If we choose $a=0$, $\delta= 1/b $\quad and\quad $G(t)$ as in (\rf{I28}), then $V_1$ is conjugate to $Y(g)$. On the other hand if we set $a\ne 0$, $\delta = 1/b $, $\ep = 1/c $ and define $F(t)$ and $G(t)$ as\\
\be
F(t)= \frac{c}{2a^2\sigma} \int \left[{{-g^2}+ g^\prime \int g(t)dt - f(t)}\right]dt + c_1, \quad
G(t)=-\frac{b}{a} \int g(t) dt + c_2, \lb{Ia29}
\ee where $c_1$ and $c_2$ are arbitrary constants, then $V_1$ is conjugate to $T_0$. If $a=g=0$ then $V_1$ is conjugate to $X(f)$.

{\bf Case 3.} \ $\beta, \gamma $ are arbitrary constants and $\sigma = e^{\lambda t}$.

Conjugating the general element $V_2 = X(f)+Y(g)+aD_\lambda , a \ne 0$ by $X(F)+Y(G)$ we obtain
\bea
&&Ad\left\{\exp(\ep X(F)+\delta Y(G))\right\}V_2 \nn\\ 
&=&V_2-\epsilon \ [X(F),aD_\lambda] -\delta [Y(G), Y(g)]- \delta [Y(G), aD_\lambda],\nn\\
&=& aD_\lambda + X(f+a\epsilon F^\prime - \frac{\delta}{2\sigma}[gG^\prime-Gg^\prime])- a\delta Y(\frac{\lambda}{2}g-g^\prime) -a \delta X(\frac{\lambda}{2}\frac{yg^\prime}{\sigma}),\nn\\
&=&aD_\lambda + X(f+a\epsilon F^\prime - \frac{\delta}{2\sigma}[gG^\prime-Gg^\prime]-\frac{a\delta\lambda}{2}\frac{yg^\prime}{\sigma})+Y(g-\frac{a\delta\lambda}{2}g+a\delta g^\prime).  \lb{I30}
\eea
If we choose  $ a \ne 0, \ep = 1/d $, $g =0$ and fix $F(t)= -\frac{d}{a}\int f(t)dt + c_1$,   
   then $V_2$ is conjugate to $D_\lambda $. If $a=0$, $G(t)$ as in (\rf{I28}),  then $V_2$ is conjugate $Y(g)$. If $a=g=0$, then $V_2$ is conjugate to $X(f)$.

{\bf Case 4.} \ $\beta, \sigma$ are arbitrary constants and $\gamma(t) = e^{\lambda t}$. 

Conjugating the general element $V_4 = X(f)+Y(g)+aE_\lambda,\ a\ne 0$ by $X(F)+Y(G)$ we obtain
\bea
&&Ad\left\{\exp(\ep X(F)+\delta Y(G))\right\}V_4 \nn\\
 &=&V_4-\epsilon \ [X(F),aE_\lambda] -\delta [Y(G), Y(g)]
- \delta [Y(G), aE_\lambda],\nn\\
&=& aE_\lambda + X(f-\frac{a\epsilon\lambda}{3} F + {a\epsilon }F^\prime-\frac{\delta}{2\sigma}(gG^\prime-Gg^\prime))+Y(g - \frac{a\delta\lambda G}{6}+\delta a G^\prime). \lb{I32}
\eea
Again we can shown that $V_4$ is conjugate to either one of the generators $X(f), Y(g), E_\lambda$.

{\bf 5. Reductions to (1+1) dimensional PDEs.}

The general method for performing the symmetry reduction using some specific subgroup $G_0$ of the symmetry group G is to first find the invariants of $G_0$ and rewrite (\rf{I2}) in terms of these invariants. The invariants are obtained by solving the system of PDEs $X_iI(x,y,t,u)=0, \ i=1,...,r,$
where ${X_1,X_2,...,X_r}$ is a basis for the Lie algebra of the symmetry group $G_0$.

{\bf 5.1 Subalgebra}\ $L_{s,1} = \left\{X(f)\right\}$.\ 
Integration of the one-dimensional vector field $X(f)$, where $f(t)$ is arbitrary leads to 
\be
u(x,y,t) = \frac{f^\prime(t)}{f(t)}x + W(y,t).\lb{I33}
\ee
Insertion of (\rf{I33}) into (\rf{I2}) yields the PDE
\be
\left(\frac{f^\prime}{f}\right)^\prime + \left(\frac{f^\prime}{f}\right)^2 + \beta \left(\frac{f^\prime}{ f}\right) +\sigma W_{yy}=0\lb{I34}
\ee
If we denote ${f^\prime}/{f}$ by $F(t)$, then equation (\rf{I34}) can be integrated to yield
\be
W(y,t)= -\frac{1}{\sigma}(F^\prime + F^2 + \beta F) \frac{y^2}{2}+h(t)y+k(t).\lb{I35}
\ee
Thus we obtain the following  family of solutions of (\rf{I2}) which involve three arbitrary functions $f(t), h(t)$ and $k(t)$ of time t, by inserting (\rf{I35}) into (\rf{I33}):
\be
u = \frac{f^\prime}{f}x   -\frac{1}{\sigma}(F^\prime + F^2 + \beta F) \frac{y^2}{2}+h(t)y+k(t).\lb{Ia35}
\ee

{\bf 5.2 Subalgebra}\ $L_{s,2} = \left\{Y(g)\right\}$\\
We use the ansatz
\be
u=W(\xi ,\eta) - \frac{y^2}{4g}\left(\frac{g^\prime}{\sigma}\right)^\prime , \quad  \xi=\frac{y^2}{2}+\frac{2g\sigma}{g^\prime}x , \quad \eta = t,\lb{I36}
\ee
into (\rf{I2}) and obtain the  PDE
\be
G^2 WW_\xi + \beta GW + \gamma G^4 W_{\xi\xi\xi}+ \sigma W+ G W_\eta+G^\prime \xi W_\xi=0,\qu G(\eta) = \frac{2 g \sigma}{g^\prime}.\lb{I37}
\ee
 
If we choose 
\be
 \sigma +\beta G=G^\prime,
  \lb{I38} 
  \ee
then (\rf{I37}) admits a first integral
\be
\left(\frac{1}{2}G^2 W^2+ G^\prime \xi W + \gamma(\eta) G^4 W_{\xi\xi}\right)_\xi+G W_\eta=0.\lb{I39}
\ee
Further if we assume that $ G=c$
where c is a constant,
then (\rf{I39}) reduces to 
\be
W_\eta + cWW_\xi + c^3\gamma(\eta)W_{\xi\xi\xi} =0.\lb{I41}
\ee
which is a variable coefficient K-dV equation. We note that a generalized version of (\rf{I41}) in the form
\be
u_t + u^n u_x + \alpha(t) u + \beta(t) u_{xxx}=0,\lb{I42}
\ee
 has recently been studied for its symmetry group and similarity solution by Senthilkumaran,  Pandiaraja and Mayil Vaganan [13]. Equation (\ref{I41}) is a special case of (\ref{I42}) if $\alpha$ is a constant.

The two conditions $G(\eta) = {2 g \sigma}/{g^\prime} $ and $G=c$ lead to the determination of $g(t)$ and $\beta(t)$ in terms of $\sigma(t)$
\be
g(t) = g_0 e^{\frac{2}{c}\int \sigma(t)} dt, \quad \beta(t)=-\frac{1}{c}\sigma(t).\lb{I43}
\ee

{\bf5.3 Subalgebra}\ $L_{s,3} = \left\{T_0\right\}$\\
The change of variables  $u=W(\xi, \eta), \ \xi = x,\  \eta =y $ replaces (\rf{I2}) by 
\be
(WW_\xi + \beta W + \gamma W_{\xi\xi\xi})_\xi + \sigma W_{\eta\eta} = 0.\lb{s1}
\ee
{\bf5.4  Subalgebra} $L_{s,4} = \left\{D_\lambda\right\}$

Insertion of  
$u=W(\xi,\eta), \xi =x,\  \eta = y e^{- \frac{\lambda}{2}t}$ into (\rf{I2}) changes the latter to
 \be
\left(-\frac{\lambda}{2}\eta W_\eta + WW_\xi + \beta W + \gamma W_{\xi\xi\xi}\right)_\xi+ W_{\eta\eta}=0.\lb{I44}
\ee

{\bf5.5  Subalgebra}\ $L_{s,5} = \left\{E_\lambda\right\}$

Under the transformation $u = e^{{\lambda t}/{3}}W(\xi,\eta),\  \xi = xe^{-{\lambda t}/{3}}, \ 
\eta = y e^{-{\lambda t}/{6}}, $
(\rf{I2}) becomes
\be
-\frac{\lambda}{3}\xi W_{\xi\xi}-\frac{\lambda}{6} \eta W_{\eta\xi} + W_\xi ^2+ WW_{\xi\xi}+\beta W_\xi+W_{\xi\xi\xi\xi}+\sigma W_{\eta\eta}=0. \lb{I47}
\ee

{\bf 6. \ Reduction to ODEs}

We shall now reduce the PDEs (\rf{s1}), (\rf{I44}), (\rf{I47}) to ODEs by imbedding 
$T_0, D_\lambda $ and $E_\lambda $ into two dimensional subalgebras of the the symmetry algebra of the GKPE. For, we commute $T_0, D_\lambda $  and $E_\lambda$ with $V = X(f)+Y(g)$ and require that they form a two-dimensional subalgebra. As a consequence, the function $f(t)$ and $g(t)$ get defined in terms of t. As there are two isomorphy classes of two-dimensional Lie algebras,namely, Abelian and non-Abelian,we shall take this fact into account in the foregoing analysis.

{\bf 6.1	Abelian Subalgebras}

{\bf 6.1.1}	Abelian Subalgebra. $L_{a,1}=\left\{T_0,X(1)+Y(1)\right\}$

Now we reduce the PDE (\rf{s1}) to an ODE by imbedding $T_0$ into two-dimensional Abelian subalgebra $L_{a,1}$ of the the symmetry algebra of the GKPE (\rf{I2}). Indeed, the transformation  $W = H(\rho), \rho= \xi - \eta$ replaces  (\rf{s1}) by  the third order ODE
\be
HH^\prime + \beta H + \gamma H^{\prime\prime\prime} - \sigma H^\prime =0.\lb{I49} 
\ee

{\bf 6.1.2}	Abelian Subalgebra $L_{a,2}=\left\{D_\lambda,X(1)\right\}$

Now we  reduce the PDE (\rf{I44}) through the transformation $W = H(\rho),\ \rho = \eta$ to   
\be
H^{\pr\pr}=0.\lb{I51} 
\ee

{\bf 6.1.3}	Abelian Subalgebra $L_{a,3}=\left\{E_\lambda,X(e^{\frac{\lambda}{3}t})+Y(e^{\frac{\lambda}{6}t}f)\right\}$

Now we reduce the PDE (\rf{I47}) to a ODE by imbedding $E_\lambda$ into two-dimensional Abelian subalgebra $L_{a,3}$ of the the symmetry algebra of the GKPE. The transformation 
\be
W = \frac{\lambda}{3} \eta - \frac{\lambda^2}{144\sigma}\eta^2+H(\rho),\quad \rho= \xi - \eta + \frac{\lambda}{24\sigma}\eta^2 \lb{I54}
\ee
reduces (\ref{I47}) to the fourth order ODE
\be
H^{iv}+ HH^{\prime\prime} +(-\frac{\lambda}{3}\rho + \sigma)H^{\prime\prime}+ {H^{\prime}}^2 +(\beta + \frac{\lambda}{12})H^{\prime}-\frac{\lambda^2}{72}=0\lb{Ia54}
\ee
Integrating (\rf{Ia54}) with respect to $\rho$, we get\be
H^{\prime\prime\prime}+HH^\prime +\sigma H^\prime -\frac{\lambda}{3}\rho H^\prime + (\beta +\frac{5\lambda}{12})H-\frac{\lambda^2 \rho}{72}=c_1\lb{I55}
\ee
Equation (\rf{I55}) can again be integrated to 
\be
 H^{\prime\prime} +\frac{H^2}{2}+\sigma H - \frac{\lambda}{3}\rho H - \frac{\lambda^2}{144}\rho^2+c_1\rho+c_2=0 ,   \qu {\rm if}\qu \beta = -\frac{3\lambda}{4}. \lb{I59} 
 \ee
 
{\bf 6.2	Non-Abelian Subalgebras}

{\bf 6.2.1}	Non Abelian Subalgebra $L_{n,1}=\left\{T_0,X(e^t)+Y(e^t)\right\}$

Now we reduce the PDE (\rf{s1}) to an ODE by imbedding $T_0$ into two dimensional non-Abelian subalgebra $L_{n,1}$ of the the symmetry algebra of the GKPE (\rf{I2}).

Invariance under the two dimensional subalgebra $L_{n,1}$  gives 
\be
W = H(\rho)+\xi,\quad \rho= \xi - \eta + \frac{\eta^2}{4\sigma}, \lb{I60}
\ee
where $H(\rho)$ satisfies the fourth  order ODE
\be
H^{(iv)}+H^{\prime\prime}\rho +\sigma H^{\prime\prime}+{H^{\prime}}^2+ HH^{\prime\prime}+\left(\beta+\frac{3}{2}\right)H^\prime+\left(1+\beta\right)=0.\lb{I61}
\ee
Integration of (\rf{I61}) results in
\be
H^{\prime\prime\prime}+H^{\prime}\rho +\sigma H^{\prime}+ HH^\prime+\left(\beta+\frac{3}{2}\right)H+(1+\beta)\rho=c_1, \lb{62}
\ee
which under the condition $\beta = -\frac{1}{2},$ changes to
\be
 H^{\prime\prime} + \rho H + \sigma H + \frac{H^2}{2}+\frac{1}{4} \rho^2+c_1 \rho +c_2=0.\lb{I62} 
\ee
{\bf 6.2.2}	Non-Abelian Subalgebra $L_{n,2}=\left\{D_\lambda,X(e^t)\right\}$

Now under $W = \xi + H(\rho),\ \rho = \eta $ changes to
$ 
H^{\prime\prime} + (1 + \beta)=0.
$

{\bf 6.2.3}	Non-Abelian Subalgebra $L_{n,3}=\left\{E_\lambda,\  X(e^{(1+\frac{\lambda}{3})t})+Y(e^{(1+\frac{\lambda}{6})t})\right\}$

Now we reduce the PDE (\rf{I47}) to a ODE by imbedding $E_\lambda$ into two-dimensional Abelian subalgebra $L_{n,3}$ of the the symmetry algebra of the GKPE (\ref{I2}). Equation (\rf{I47}), under the similarity transformation
\be
W = \frac{3+\lambda}{3} \eta - \frac{6+\lambda^2}{144\sigma}\eta^2+H(\rho),\quad \rho= \xi - \eta + \frac{6+\lambda}{24\sigma}\eta^2, \lb{II70}
\ee
reduces to
\be
 H^{\prime\prime\prime} +HH^{\prime}-\frac{\lambda}{3}\rho H^{\prime}+H^{\prime} + (\beta+\frac{\lambda}{3})H- \frac{(6+\lambda)^2}{72}\rho=c_1,\lb{I72} 
\ee
If $\beta =-\frac{2\lambda}{3},$ then (\rf{I72}) can be integrated to yield 
\be
H^{\prime\prime} +\frac{\lambda}{2} H^2-\frac{\lambda}{3}\rho H+H - \frac{(6+\lambda)^2}{144} \rho^2+c_1\rho+c_2=0.\lb{I73}
\ee

{\bf 7. The general form of reductions of GKPE (\ref{I2}) }

The transformation $H(\rho) = f^{-1}(\rho)$ replaces the ODEs (\ref{I59}), (\ref{I62}) and (\ref{I73}), respectively, by  
\bea
ff^{\prime\prime} -2{f^{\prime}}^2 - \frac{1}{2}f - \sigma f^2 + \frac{\lambda}{3}\rho f^2 + \left( \frac{\lambda ^2}{144} \rho ^2 -c_1 \rho -c_2 \right)f^3&=&0, \lb{aa}\\
 ff^{\prime\prime} -2{f^{\prime}}^2 - \frac{1}{2}f - \left(\rho + \sigma \right) f^2  + \left( \frac{1}{4} \rho ^2 +c_1 \rho +c_2 \right)f^3&=&0 \lb{bb}\\
ff^{\prime\prime} -2{f^{\prime}}^2 - \frac{\lambda}{2}f - \left(\frac{\lambda}{3}\rho -1\right) f^2  + \left( \frac{(\lambda +6)^2}{144} -c_1 \rho-c_2  \right)f^3&=&0. \lb{cc}
\eea
We may write the general form of the equations (\rf{aa}), (\rf{bb}), (\rf{cc}) as
\be
ff^{\prime\prime}+a{f^{\prime}}^2 + bf + g(\rho)f^2+ h(\rho)f^3=0.\lb{dd}
\ee
which is a special case of the equation introduced by Mayil Vaganan and Senthilkumaran [11], viz.,
\be
ff^{\prime\prime} + a(\rho){f^{\prime}}^2 + b(\rho)ff^{\prime} +c(\rho)f^2 + d(\rho)f^{\prime} + g(\rho)f^3 +k f=0. \lb{ff}
\ee

{\bf 8.Conclusions}

We now summarize the results of the present work,below:

\quad As emphasized by David, Karman, Levi and Winternitz [1] and G\"ung\"or [2] that it is of great interest to identify all nonolinear PDEs that admit infinite-dimensional symmetry groups and Lie algebras containing arbitrary functions.

\quad In this paper we have shown that the GKPE (\rf{I2})is one such equation. When all the four functions $\beta(t), \gamma(t)$  and $\sigma(t)$ are kept arbitrary. The GKPE (\rf{I2}) is shown to admit an infinite-dimensional symmetry group with a Lie algebra $L_p$ involving two arbitraray functions $f(t)$ and $g(t)$. Further we extend the Lie algebra $L_p$ into four Lie algebras $L_i, i=1,2,3,4$ by taking 
$\sigma, \gamma$ to be equal to $e^{\lambda t}$.

\quad The classification of  one-dimensional subalgebras of the symmetry algebra under the adjoint action of the symmetry group is carried out. Then by commuting $T_0, D_\lambda, E_\lambda$ with $V=X(f)+Y(f)$ 
two-dimensional subalgebras are constructed.

\quad The GKPE (\rf{I2}) is also shown to reduce to a linear PDE of the form $W_{yy}=F(f(t),f^\prime(t))$ (cf.(\rf{I51})),  a variable coefficient-KdV equation (\rf{I41}).

\quad The reduction of the GKPE (\rf{I2}) into ODEs (\rf{aa}), (\rf{bb}), (\rf{cc}) under Abelian 
subalgebras and non-Abelian subalgebras are of the form (\rf{dd}).  

We also have found a new solution (\ref{Ia35}) of (\rf{I2}) involving two arbitrary functions.
 
A rigorous analysis of the equation (\ref{dd}) or its generalized version (\ref{ff}) is yet to be studied.

\bc
{\bf References} 
\ec
\begin{enumerate}
\item Bluman G W and Kumei S \ 1989 \ {\it Symmetries and Differential Equations}\  (New York: Springer-Verlag)
\item Brugarino T and Greco A M \ 1991\ {\it J. Math. Phys.}\ {\bf 31}\ 69-71
\item David D, Kamran N D, Levi D and Winternitz P\ 1986\   {\it J. Math. Phys.}\ {\bf27}\ 1225-1236 
\item David D, Levi D and Winternitz P\ 1987\   {\it Stud. Appl. Math.,}\ {\bf76}\ 133-138  
\item David D, Levi D and Winternitz P\ 1989\   {\it Stud. Appl. Math.,}\ {\bf80}\ 1-23 
\item Gerd Baumann\ 2000\ {\it Symmetry Analysis of Differential Equations with Mathematica}\ (New York: Springer-Verlag)
\item Gungor F \ 2001\ {\it J. Phys. A: Math. Gen.}\ {\bf 34}\ 4313-4321
\item Gungor F \ 2001\ {\it J. Phys. A: Math. Gen.}\ {\bf 35}\ 1805-1806
\item Gungor F and Winternitz P\ 2002\  {\it  J. Math. Anal. Appl.}\ {\bf  276}\ 314-328 
\item Gungor F and Winternitz P\ 2004\  {\it  Nonlinear Dynamics}\ {\bf  35}\ 381-396 
\item Mayil Vaganan B and Senthil Kumaran M\ 2008\		{\it Nonlinear Analysis: Real World Applications,} \ {\bf 9}\ 2222-2233
\item Olver P J \ 1986 \ {\it Applications of Lie Groups to differential
 equations}\  (New York: Springer-Verlag)
\item Senthilkumaran M, Pandiaraja D and Mayil Vaganan B\ 2008 \ {\it Appl. Math. comp.}\ {\bf 202}\ 693-699 
\item Yao Ruo-Xia and Lou Sen-Yue\ 2008\    {\it Chin. Phys. Lett}\ {\bf 25}\ 1927-1931
\end{enumerate}

\end{document}